# Comprehensive performance comparison among different types of features in data-driven battery state of health estimation

Xinhong Feng [1], Yongzhi Zhang [1,*], Rui Xiong [2,*], Chun Wang[3]

[1] College of Mechanical and Vehicle Engineering, Chongqing University, 400030, Chongqing, China.
[2] Department of Vehicle Engineering, School of Mechanical Engineering, Beijing Institute of Technology, Beijing 100081, China
[3] School of Mechanical Engineering, Sichuan University of Science and Engineering, Sichuan, 643000, China,
[*] **Corresponding authors**: Yongzhi Zhang (yongzhi@cqu.edu.cn), Rui Xiong (rxiong@bit.edu.cn)

**Abstract:**

Battery state of health (SOH), which informs the maximal available capacity of battery, is a key indicator of battery aging failure. Accurately estimating battery SOH is a vital function of the battery management system that remains to be addressed. In this study, a physics-informed Gaussian process regression (GPR) model is developed for battery SOH estimation, with the performance being systematically compared with that of different types of features and machine learning (ML) methods. The method performance is validated based on 58826 cycling data units of 118 cells. Experimental results show that the physics-driven ML generally estimates more accurate SOH than other non-physical features under different scenarios. The physical features-based GPR predicts battery SOH with the errors being less than 1.1% based on 10 to 20 mins' relaxation data. And the high robustness and generalization capability of the methodology is also validated against different ratios of training and test data under unseen conditions. Results also highlight the more effective capability of knowledge transfer between different types of batteries with the physical features and GPR. This study demonstrates the excellence of physical features in indicating the state evolution of complex systems, and the improved indication performance of these features by combing a suitable ML method.

## I. INTRODUCTION

Lithium-ion (li-ion) batteries have been widely used in electric vehicles (EVs), stationary energy storage, and portable electronics including smartphones, laptops, *etc.* [1]-[3]. Despite of the high energy density and long cycle life, li-ion batteries degrade in usage, and it is important to monitor the battery aging states online for safety concerns. The state of health (SOH), as a key indicator of battery aging evolution, informs the maximum available capacity of battery and the battery aging failure [4]-[6]. Therefore, it is a vital function for the battery management system (BMS) to accurately estimate battery SOH. Although a lot of efforts have been paid to estimating accurate battery SOH from both academia and industry, it is a non-trivial problem that remains to be addressed [7], [8].

The online SOH estimation of battery is mainly conducted based on two types of methods, including model-based and data-driven methods. Model-based methods usually construct state-space equations based on equivalent circuit models (ECMs) [9], [10] or electrochemical models [11]-[14], with SOH as a state value to be estimated. ECMs are formulated with simple mathematics and thus easy to be deployed onboard with low computational burden. However, ECMs are designed to mimic the battery charging/discharging behaviors and cannot be used to capture the variations of battery internal states and parameters. And the modeling accuracy of ECMs varies as the ambient temperatures change and deteriorates as battery degrades. To improve the estimation accuracy and robustness, the electrochemical models, which can capture the varying trajectories of battery internal states and parameters, are developed to describe battery physical characteristics. And based on these models, a high estimation accuracy of battery SOH can be kept at different temperatures and aging states [11], [12]. Furthermore, some interesting and informative parameters of battery aging such as the thickness of solid electrolyte interphase, which cannot be predicted by ECMs, can however, be predicted using electrochemical models [13], [14].

Although showing much promise in estimating battery SOH, the electrochemical modeling still suffers from several issues. First, a full electrochemical model describing battery aging couples several sub-models that simulate different aging mechanisms. The research on these aging mechanisms remains on a shallow level, and more effort is required for deeper insights [15], [16]. Second, the electrochemical models possess tens of vital parameters to be identified, and the parameter identification process involves in tedious disruptive and offline experiments, leading to a difficult task for quick and accurate identification of electrochemical parameters [17]. Finally, unlike the simple mathematics of ECMs, the electrochemical model solving involves in intensive calculations of partial differential equations, which poses a high requirement for the hardware when deploying onboard [18].

Recently, as the fast development of artificial intelligence (AI) and machine learning (ML) theory, more and more researchers devote to developing data-driven techniques for online battery SOH estimation. For data-driven methods, a large amount of training data is typically required to construct a ML model. To improve the prediction accuracy, researchers have been spending abundant time on extracting a variety of features from battery measurement data (*e.g.*, voltage, current, and temperature). Typically, these features can be classified into three types including physical features, statistical features, and original features. The physical features contain rich information of battery aging evolution, and the prediction results can thus be reasonably interpreted [19], [20]. The statistical features extract statistics results, *i.e.*, the maxima, mean, variance, skewness, *etc.*, from the collected data [21], [22]. And the dimensionality of the input features is thus reduced. The original features are directly sampled from the measurement data of current, voltage and temperature, or the transformed results of the data, *i.e.*, incremental capacity curve and differential voltage curve [23], [24]. In this case, the input dimensions are generally high since enough data points need to be used to capture the curve changing characteristics.

Normally, different features are fused together to obtain the most accurate SOH estimation in the literature. For example, Paulson et al. [25] extract a total of 396 features (physical and statistical) from the battery measurement data, and the features most related to battery aging is further selected based on the Pearson correlation coefficient. Zhang et al. [26] generate 206 features of three types for feature selection, and only 7 features are selected for tracking SOH change based on an automatic feature selection pipeline. By combining four algorithms, Xiong et al. [27] develop a feature selection method to select the most relevant features, from a total of 568 features of three types, to indicate battery aging. However, a systematical performance comparison between these features are missing in the literature. With feature engineering, a great number of ML methods, including Gaussian process regression (GPR), deep neural networks (DNN), support vector regression (SVR), *etc.*, have been developed to construct the relationships between the extracted features and battery SOH. Indeed, the estimation accuracy based on the features can be improved by combining with proper ML models [25]-[27].

In this study, we have made three contributions to advance the current battery SOH estimation technique. First, six physical features of an equivalent circuit model are extracted from the voltage relaxation data that is easily collected onboard. The GPR model is constructed to learn the relationships between these features and battery SOH, with the model performance systematically compared with that of different ML methods combined with different features, *i.e.*, statistical and original features. Second, the performance of these models, *e.g.*, estimation accuracy and generalization capability, is validated against different ratios of training and unseen test data. And the transfer learning performance between different battery types is also evaluated. Finally, the experimental data of three types of batteries, containing 58826 cycling data units of 118 cells, are used to verify the methods. Results show that in most cases, the physical features perform better in battery SOH estimation than other features. And the health indication capability of physical features is further improved based on GPR, indicating the necessity of combining the features with a suitable ML method for highly accurate SOH estimation.

The paper is organized as follows. Section 2 introduces the experimental data. Section 3 presents the features extraction and ML methods, followed by Section 4 showing the features extraction and battery SOH estimation results. Section 5 concludes the paper.

## II. DATA GENERATION

The dataset used in this study, which was generated in ref. [22], include cycling data of three types of battery cells. According to the cathode materials of cells, the dataset are divided into three sub-datasets (Datasets 1, 2 and 3). Datasets 1, 2 and 3 were, respectively, generated from battery cells with the cathode materials mainly consisting of $Li_{0.86}Ni_{0.86}Co_{0.11}Al_{0.03}O_2$ (NCA), $Li_{0.84}(Ni_{0.83}Co_{0.11}Mn_{0.07})O_2$ (NCM), and a blend of $Li(NiCoMn)O_2$ - $Li(NiCoAl)O_2$ (NCM+NCA). The anode of NCA and NCM cells contains about 97 wt% graphite and 2 wt% silicon, while the anode of NCM+NCA cell mainly consists of graphite. The nominal capacity of NCA and NCM cells is 3.54 Ah, and the NCM+NCA cell has a nominal capacity of 2.5 Ah. The lower cutoff voltage of the NCA battery is 2.65 V, which value is 2.5 V for the NCM and NCM+NCA batteries. The upper cutoff voltage of all battery cells is 4.2V. The experimental temperature is controlled through climate chambers, with three different temperatures set in the aging experiment, including 25℃, 35℃, and 45℃.

Table I lists the detailed cycling conditions for each type of battery. The charge current rate is 0.5 C in most cases except one case of the NCA battery under 25℃, when the charge current rate is 0.25 C. The discharge current rate of the NCA and NCM battery is 1 C, while for the NCM+NCA battery, the discharge rates range from 1 C to 4 C. For each charge-discharge cycle, the cell is charged/discharged with a constant current (CC) to the corresponding cutoff voltages, and during the charging process a constant-voltage (CV) charge is also loaded with a cutoff current of 0.05C. After the completion of each charge or discharge, the NCA and NCM cells are rested for 30 minutes while the NCM+NCA cells are rested for 60 minutes. Table I also shows the number of cells and data points under each cycling condition. Totally, there are 118 cells with 58826 data units used to validate the developed algorithms in this study. Note that for simplicity, the cell under each cycling condition is named CYA-B/C, where A represents the experimental temperature, and B and C denote the charge and discharge current rates, respectively.

Table I Cycling conditions for the dataset generation.

| Datasets | Temperature (±0.2℃) | Current (C-rate) | | Number of cells | Number of data units |
|---|---|---|---|---|---|
| | | Charge | Discharge | | |
| Dataset 1 | 25 | 0.25 | 1 | 6 | 1761 |
| | 25 | 0.5 | 1 | 19 | 3290 |
| | 35 | 0.5 | 1 | 2 | 1123 |
| | 45 | 0.5 | 1 | 28 | 15928 |
| Dataset 2 | 25 | 0.5 | 1 | 22 | 5488 |
| | 35 | 0.5 | 1 | 4 | 4806 |
| | 45 | 0.5 | 1 | 28 | 17666 |
| Dataset 3 | 25 | 0.5 | 1 | 3 | 2900 |
| | 25 | 0.5 | 2 | 3 | 2978 |
| | 25 | 0.5 | 4 | 3 | 2886 |

Fig. 1 illustrates the battery SOH degradation trajectories during cycling. Fig. 1(a) indicates that the NCA

battery deteriorates faster under lower temperatures and higher charging currents, while Fig. 1(b) shows that the optimal operation temperature for the NCM battery is 35°C, and a lower (25°C) or higher (45°C) temperature accelerates battery aging. For the NCM+NCA cell, high conformity of the capacity decay under different discharge rates is observed in Fig. (c), where the results indicate that the increase of discharge current rates does not necessarily accelerate battery aging.

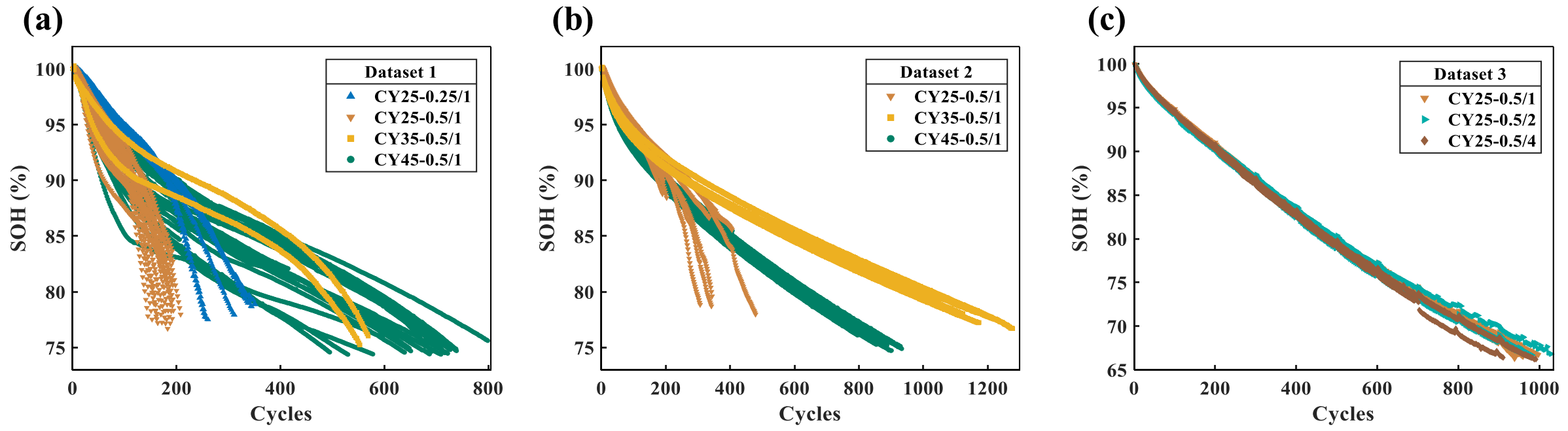

Fig. 1. Battery SOH versus cycle number of (a) the NCA battery (Dataset 1),(b) the NCM battery (Dataset 2), and (c) the NCM+NCA battery (Dataset 3).

## III. METHODOLOGY

### *A. Features Construction*

For fair comparisons, the different features used in this study are all extracted from the voltage relaxation curve after battery full charging, as shown in Fig. 2. The first type of features are the sampled voltage data points, which features are regarded as the original features (namely ORIGI). Based on ORIGI, two other types of features including statistical features (namely STATS) and physical features are, respectively, further extracted (Fig. 2). Aligned with ref. [22], STATS includes six statistical features, *i.e.,* the maxima, mean, minima, variance, skewness, and the excess kurtosis of the sampling voltages. Please refer to Appendix for the detailed expressions of these

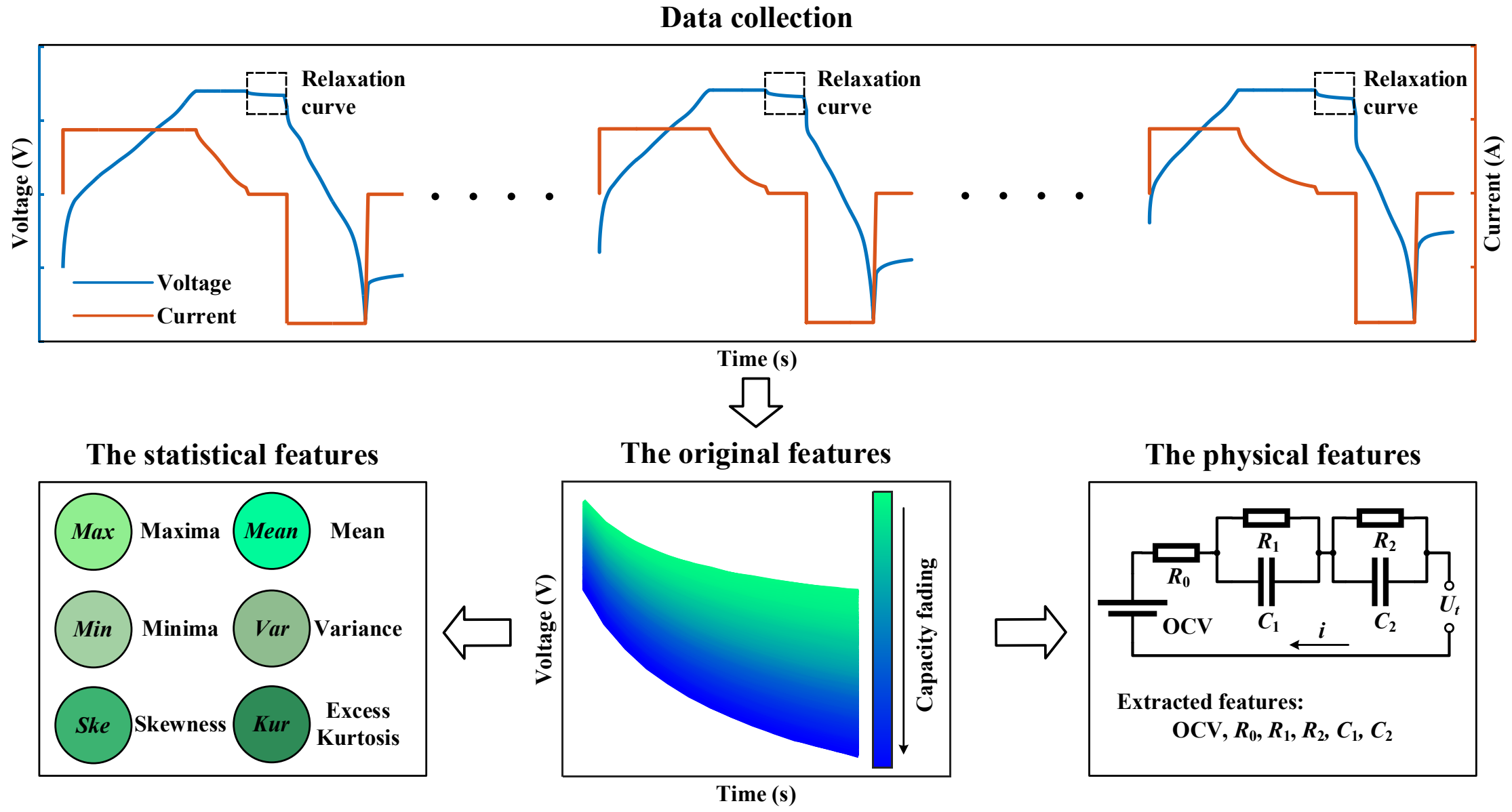

Fig. 2. Different types of features extraction based on the voltage relaxation curve.

statistical features. Note that STATS represents one of the most advanced health indicators in the literature. The physical features used in this study are the six parameters of a second-order RC model as shown in Fig. 2, including

Note that during the battery resting period after a full charging, the data sampling interval is set to 2 min for the NCA and NCM cells, and 30 s for the NCM+NCA cells. Therefore, there are a total of 14 data points sampled every cycle for the NCA and NCM cells while a total of 119 data points sampled every cycle for the NCM+NCA cells. The parameter identification processes of the equivalent circuit model (ECM) are presented as the following. The battery terminal voltage $U_t(t)$ is calculated as:

$$U_t(t) = OCV - i(t)R_0 - U_1(t) - U_2(t), \tag{1}$$

where $i(t)$ is the discharge current (negative in charging) at moment $t$, $U_1(t)$ and $U_2(t)$ represent the voltages across $R_1$ and $R_2$ at moment $t$, respectively. During the relaxation period after battery full charging, $U_1(t)$ and $U_2(t)$ are expressed as:

$$U_1(t) = IR_1 e^{-\frac{t}{R_1C_1}}, \tag{2}$$

$$U_2(t) = IR_2 e^{-\frac{t}{R_2C_2}}, \tag{3}$$

where $I = i(0)$ represents the cutoff current of CV charging. By combining eqs. (1-3), the terminal voltage $U_t(t)$ during battery resting is derived as:

$$U_t(t) = OCV - IR_1 e^{-\frac{t}{R_1C_1}} - IR_2 e^{-\frac{t}{R_2C_2}}. \tag{4}$$

Based on eq. (4), a nonlinear least squares method was used to fit the sampling data to identify parameters of $OCV$, $R_1$, $R_2$, $C_1$, and $C_2$. The internal resistance $R_0$ is subsequently obtained by the following derivation:

$$R_0 = \frac{|U_t(0) - \mathrm{OCV}|}{I} - R_1 - R_2. \tag{5}$$

*B. Machine Learning*

Gaussian process regression (GPR) is used to learn the relationship between the extracted features and battery SOH. The GPR model is trained based on a dataset $\mathcal{D} = \{\boldsymbol{x_i}, y_i\}(i = 1, \cdots, n)$, where $\boldsymbol{X} = [\boldsymbol{x_1}, \cdots, \boldsymbol{x_n}]^T$ and $\boldsymbol{Y} = [y_1, \cdots, y_n]$ are the input and output of the training set. It is assumed that $y_i = f(\boldsymbol{x_i}) + \varepsilon_i$ ($\varepsilon_i \sim N(0, \sigma^2)$). The outputs $\boldsymbol{F} = [f(\boldsymbol{x_1}), \cdots, f(\boldsymbol{x_n})]^T$ obeys a joint Gaussian distribution with a mean function $\boldsymbol{m}(\boldsymbol{X})$ (normally assumed to be **0**) and a covariance kernel function $\boldsymbol{K}$, where $\boldsymbol{K_{ij}} = k(\boldsymbol{x_i}, \boldsymbol{x_j})$. The kernel measures the similarity of the points $\boldsymbol{x_i}$ and $\boldsymbol{x_j}$.

Given an input $\boldsymbol{x}^*$, the joint probability distribution of the training set and the predicted output is expressed as:

$$\begin{bmatrix} \boldsymbol{Y} \\ y^* \end{bmatrix} \sim GP\left(\begin{bmatrix} \boldsymbol{m}(\boldsymbol{X}) \\ m(\boldsymbol{x}^*) \end{bmatrix}, \begin{bmatrix} \boldsymbol{K}(\boldsymbol{X},\boldsymbol{X}) + \sigma^2\boldsymbol{I} & \boldsymbol{K}(\boldsymbol{X},\boldsymbol{x}^*) \\ \boldsymbol{K}(\boldsymbol{x}^*,\boldsymbol{X}) & k(\boldsymbol{x}^*,\boldsymbol{x}^*) \end{bmatrix}\right), \tag{9}$$

where $y^*$ is the output to be predicted, and $\boldsymbol{I}$ is an identity matrix. The mean value ($m^*$) and variance ($\sigma^{*2}$) of $y^*$ are obtained by addressing a conditional probability distribution on the training set:

$$m^* = m(\boldsymbol{x}^*) + \boldsymbol{K}(\boldsymbol{x}^*,\boldsymbol{X})[\boldsymbol{K}(\boldsymbol{X},\boldsymbol{X}) + \sigma^2\boldsymbol{I}]^{-1}[\boldsymbol{Y} - \boldsymbol{m}(\boldsymbol{X})], \tag{10}$$

$$\sigma^{*2} = k(\boldsymbol{x}^*,\boldsymbol{x}^*) - \boldsymbol{K}(\boldsymbol{x}^*,\boldsymbol{X})[\boldsymbol{K}(\boldsymbol{X},\boldsymbol{X}) + \sigma^2\boldsymbol{I}]^{-1}K(\boldsymbol{X},\boldsymbol{x}^*), \tag{11}$$

In the study, an automatic relevance determination (ARD) exponential kernel is used for GPR, which is expressed as:

$$k_{EXP}^{ARD}(\boldsymbol{x_i}, \boldsymbol{x_j}) = \sigma_f^2 exp\left(-\sqrt{\sum_{m=1}^{d} \frac{(x_{im} - x_{jm})^2}{l_m^2}}\right), \tag{12}$$

where $l_m$ represents the length scale for the $m$th feature and $\sigma_f$ is the signal standard deviation. The hyperparameters $\theta = \{\sigma, \sigma_f, l_1, \cdots, l_d\}$ are obtained by minimizing the negative log marginal likelihood $NLML =$

$-\log P(\boldsymbol{Y}|\boldsymbol{X},\theta)$.

Note that another two advanced machine learning methods including the extreme gradient boosting (XGBoost) and support vector machine regression (SVR) are also used for comparisons in this study. And the mathematical modeling of these two methods is referred to refs. [28, 29]. The root mean squares error (RMSE), which is defined as the following, is used to evaluate the model performance:

$$RMSE = \sqrt{\frac{1}{n}\sum_{i=1}^{n}(y_i - \hat{y}_i)^2}, \quad (13)$$

where $y_i$ is the observation, $\hat{y}_i$ is the estimation, and $n$ is the total number of test samples.

*C. Transfer Learning*

Transfer learning (TL) is used to improve the prediction accuracy when only limited target domain (TD) data is available for training. Fig. 3 illustrates three TL methods used in this study, namely TL1 (Fig. 3(a)), TL2 (Fig. 3(b)), and TL3 (Fig. 3(c)). The source domain (SD) data contains a large amount of training data generated from different types of batteries. In this paper for example, the rich SD data is from Dataset 1 while the limited TD data is from Dataset 2 or Dataset 3. The purpose of TL is to transfer the sophisticated knowledge learned from SD data to the effective mining of limited TD data, so that the generalization capability of machine learning (ML) is further improved.

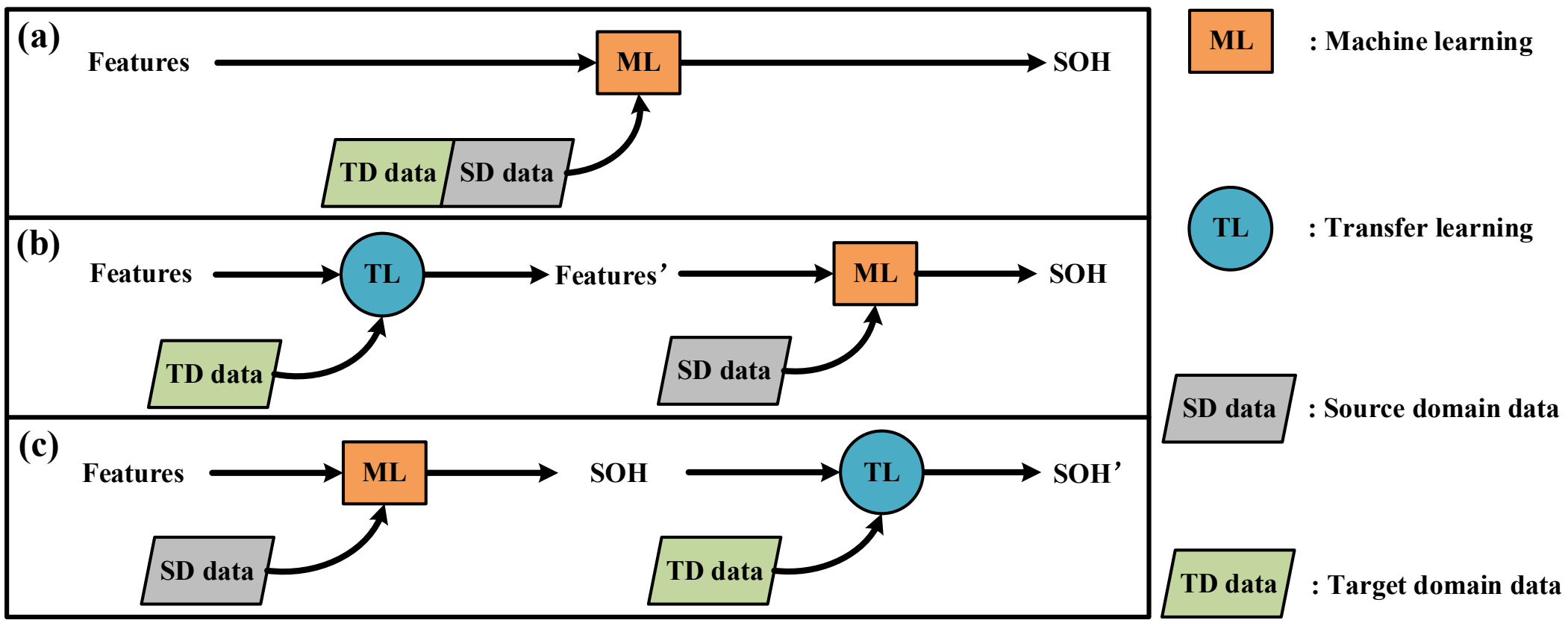

Fig. 3. Schematic of different transfer learning methods. (a) Augmentation (TL1). (b) Features transfer (TL2). (c) Delta learning (TL3).

TL1 presented in Fig. 3(a) is called 'Augmentation', where the training data based on SD is augmented by introducing some limited TD data directly. In this case, the SD and TD data are combined to train a ML model together, which model is then used to predict the target battery SOH. Fig. 3(b) illustrates a feature transfer-based TL method. In this scenario, a liner regression model is used to transform the features that are input to the ML model (trained on SD data) to minimize the prediction errors. The linear regression model is expressed as:

$$x_{ij}' = w_j x_{ij} + b_j, \quad (14)$$

where $x_{ij}$ is the *j*th feature of the *i*th sample, $x_{ij}'$ is the transformed feature, and $w_j$ and $b_j$ are coefficients that are obtained by minimizing the RMSE of SOH estimations based on TD data. TL3 shown in Fig. 3(c) shows a delta learning method, where TD data is used to train a correction model for eliminating the estimated deviations from the ML model (trained on SD data). In the case, the correction model is constructed using the same learning algorithms as the ML model. When predicting battery SOH, the target features are, respectively, input to ML to estimate the baseline SOH and input to the correction model to estimate the delta SOH. And the sum of these two estimated SOH values determines the final SOH prediction output.

## IV. RESULTS

### A. Features Extraction

Fig. 4 shows the ECM features extracted from the voltage relaxation curves at different battery aging states, where Figs. 4(a-f), respectively, shows the variations of battery OCV, $R_0$, $R_1$, $R_2$, $C_1$ and $C_2$ versus SOH of all three types of batteries. It is observed that on one hand, the features of OCV, $C_1$ and $C_2$ decrease while $R_0$, $R_1$ and $R_2$ increase monotonically as battery ages. These monotonical relationships between the features and battery SOH lay a solid foundation for accurate SOH estimation by implementing ML. On the other hand, owing to the cell-to-cell heterogeneity and distinct loading conditions, the features generally vary vastly vs. battery SOH within each dataset (especially in Datasets 1 and 2). This vast variations of features at similar battery aging states compromise the ML-based prediction performance. And it is necessary to include all six physical features as input to ML for improved prediction accuracy. The features of Dataset 1 are closer to those of dataset 2 than those of Dataset 3 in most cases, making the TL from Dataset 1 to Dataset 2 more easily, as the results shown in the TL section. The six statistical features of these three datasets, which are not presented in this paper, are referred to ref. [24] for the details.

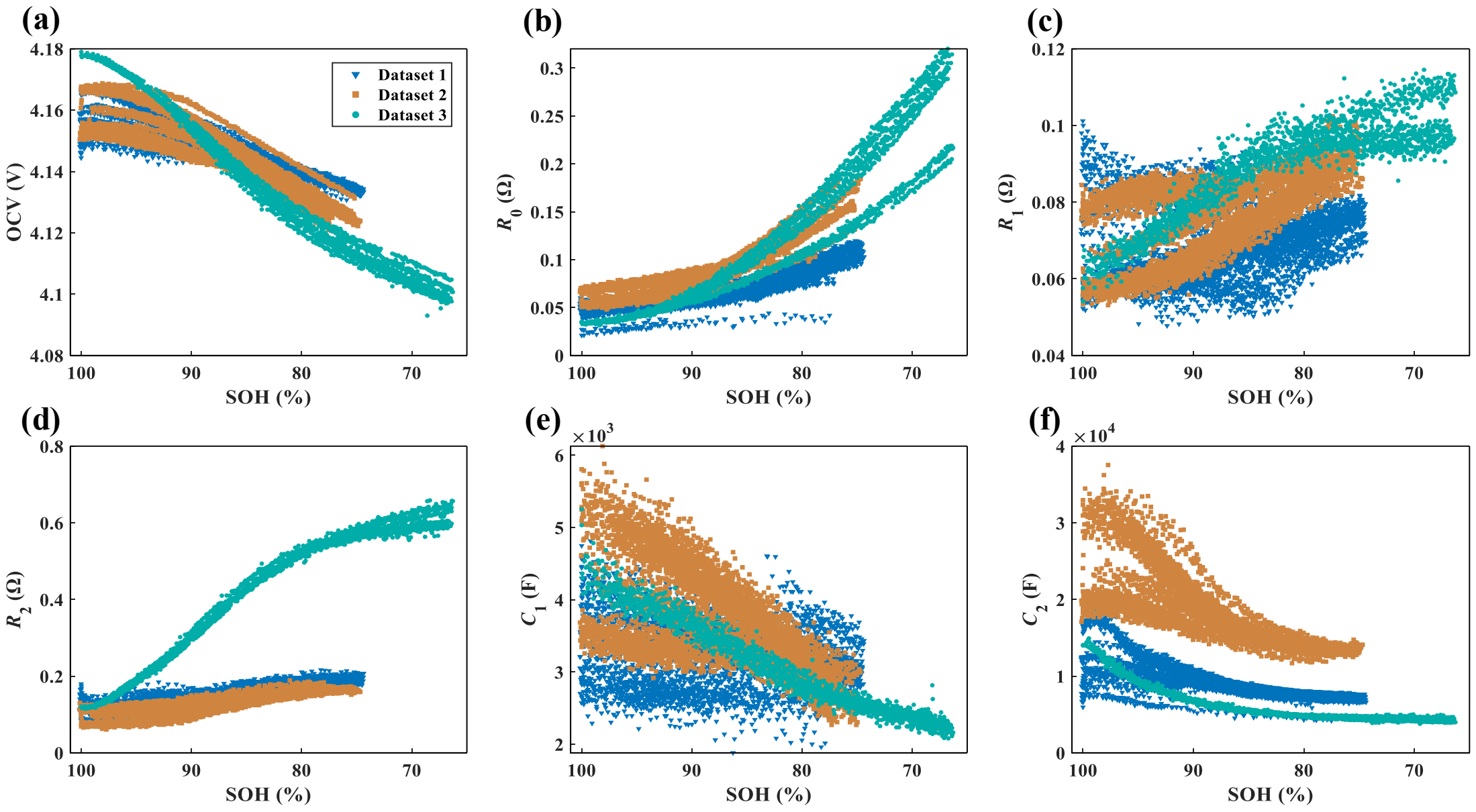

Fig. 4. ECM-based features versus battery SOH. (a) OCV. (b) $R_0$. (c) $R_1$. (d) $R_2$. (e) $C_1$. (f) $C_2$.

### B. Battery SOH estimation within the same dataset

In this section, without specific notification, cells in Datasets 1 and 2 under each operation condition are evenly divided into two groups, with one group of data used for training while another group of data used for test purposes. In Dataset 3, the data of two cells with one from the cycling condition CY25-0.5/1 and another one from CY25-0.5/4 are used for training, while the data of remaining cells are used for test.

**SOH estimation with long duration of relaxation data.** In this case, all voltage relaxation data after battery full charging is used to extract the health indicators (features). That is, in Datasets 1 and 2 a 30-min duration of voltage relaxation Data while in Dataset 3 a 60-min duration of voltage relaxation data are, respectively, used for features extraction. Fig. 5 shows the SOH estimation results using GPR, where Figs. 5(a-c) present the results based on ECM features while Fig. 5(d) presents the results based on all three types of features. Figs. 5(a-c) show that the estimated SOH is close to the observed values with the absolute estimation errors within 2% in most cases for all lower than those based on other two types of features. In Dataset 3, however, GPR estimates battery SOH with similar accuracies among different features, with the ECM-based estimation errors slightly larger than those based on other features. In general, compared with other features, the ECM-based features not only contain physical information but also indicate more accurate and reliable state about battery aging evolution.

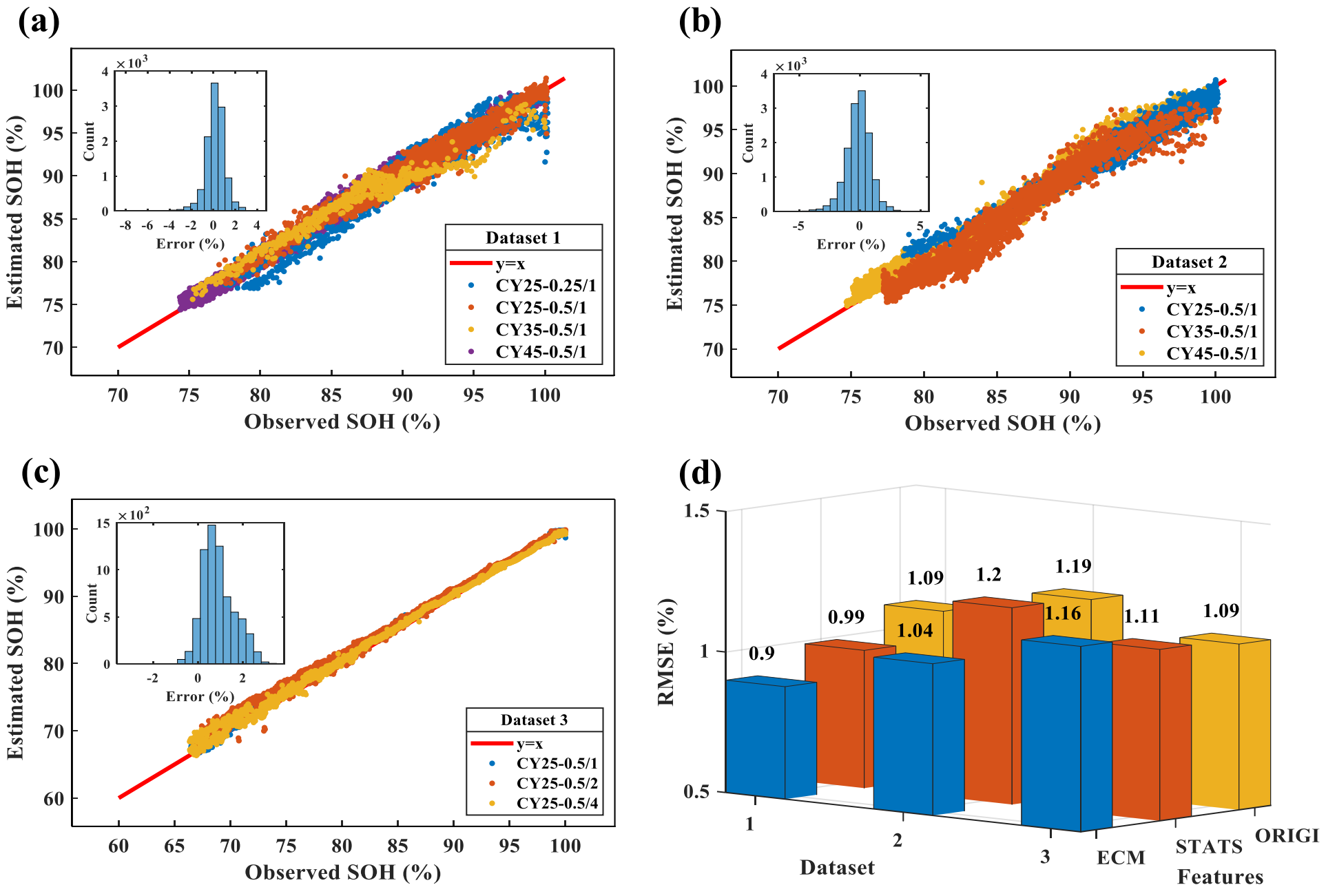

Fig. 5. SOH estimation results based on ECM features using GPR on (a) Dataset 1, (b) Dataset 2, (c) Dataset 3. (d) SOH estimation errors (RMSE) based on different types of features.

Table II lists the battery SOH estimation results achieved by different ML models, including GPR, XGBoost and SVR. The bold fonts indicate the smallest errors obtained by each method with different datasets. Generally, the ECM-based SOH estimation errors are the lowest based on different ML methods with different datasets. Only two exceptions exit with Dataset 3 when GPR and XGBoost methods are used, under which cases larger estimation errors based on ECM are obtained than those based on other features. However, SVR estimates the most accurate SOH based on ECM than other features even with Dataset 3. These highly accurate estimation results based on ECM against different ML methods again highlights the advancement of using physical features to accurately indicate battery aging states. And the GPR based on ECM generally achieves the most accurate SOH estimation among different datasets. Furthermore, the STATS-based estimation errors are generally lower than those based on ORIGI features with different ML methods and datasets, indicating the necessity to extract versatile features rather than to using the original data to estimate battery SOH.

Table II Battery SOH estimation errors (RMSE) based on different ML methods and features.

| Dataset | GPR | | | XGBoost | | | SVR | | |
|---|---|---|---|---|---|---|---|---|---|
| | ECM | STATS | ORIGI | ECM | STATS | ORIGI | ECM | STATS | ORIGI |
| Dataset 1 | **0.90** | 0.99 | 1.09 | **1.02** | 1.12 | 1.50 | **1.00** | 1.11 | 1.50 |
| Dataset 2 | **1.04** | 1.20 | 1.19 | **1.19** | 1.32 | 1.48 | **1.24** | 1.31 | 2.04 |
| Dataset 3 | 1.16 | 1.11 | **1.09** | 1.24 | 1.19 | **1.10** | **0.88** | 1.00 | 1.03 |

The SOH estimation performance based on different features is further validated under four training/test data splitting strategies, as illustrated in Fig. 6. Strategies 1 and 2 are designed to validate the methodology performance against different ratios of training and test data (*e.g.*, 1:1 and 1:2). The difference between strategies 1 and 2 is that strategy 1 distinguishes different cycling conditions while strategy 2 does not. That is, when designing strategy 1, the ratios of training and test data hold for cells cycled under the same experimental conditions, while for strategy 2, these ratios only hold for the whole dataset. Strategies 3 and 4 are designed to test the method generalizability to different usage conditions. For example, under strategy 3 the SOH at one temperature is estimated by using the ML trained on data at other two temperatures. And under strategy 4, only the first 80% amount of data of each cell is used for ML training while the rest 20% of data is used for test. Note that under strategies 1 and 2, the training/test data splitting for each case is randomly conducted by 20 times, and for each time, there will be a ML model trained based on the training data. The SOH estimation value presented in Fig. 6 is the averaged prediction of the 20 trained ML models in that case. The four strategies are mainly applied to Datasets 1 and 2 since the NCM+NCA cells (Dataset 3) show greatly high conformity against different cycling conditions, and the estimation results under these strategies may not be representative for practical applications.

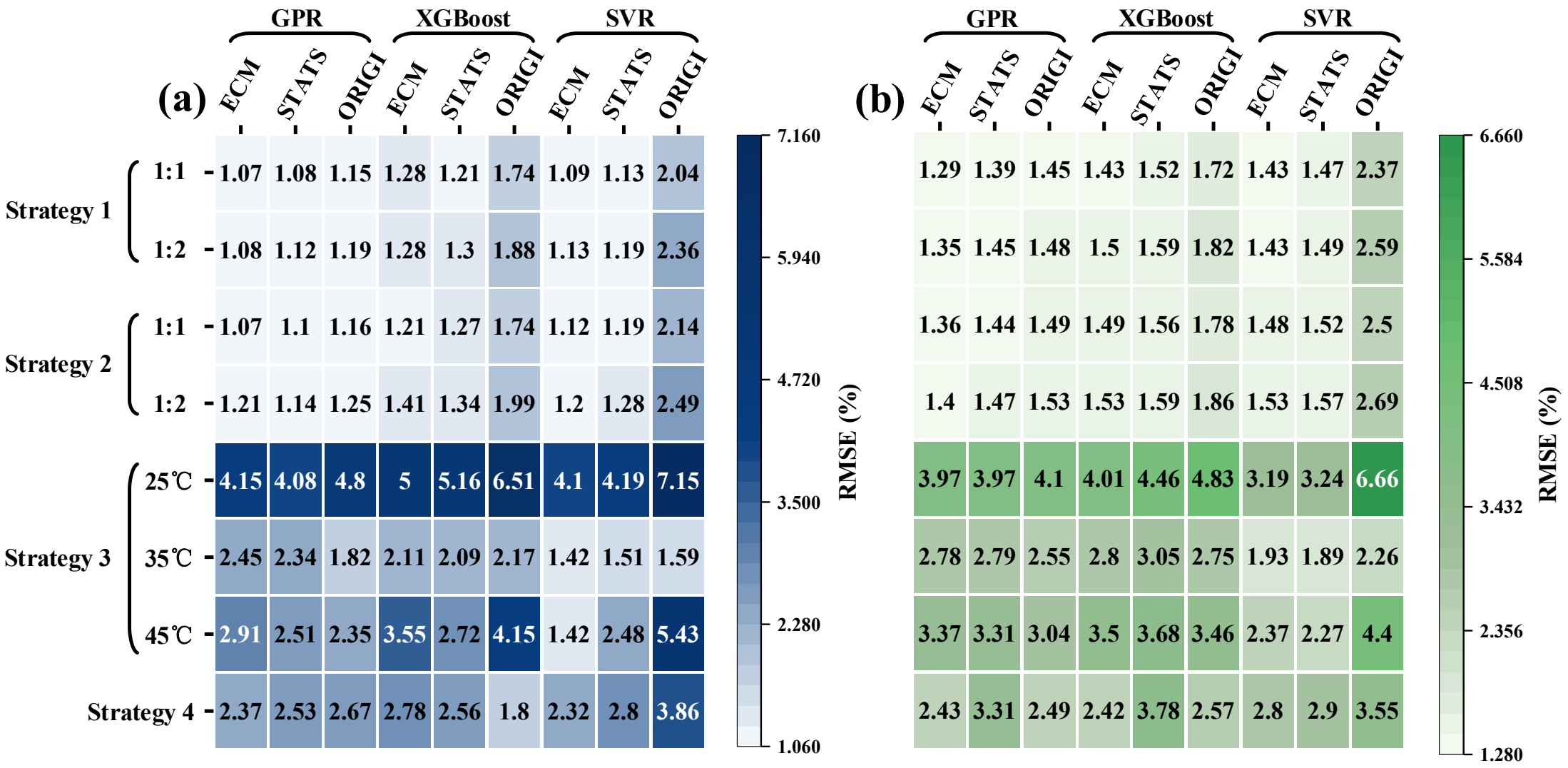

Fig. 6. RMSE of SOH estimation under the four training/test data splitting strategies. Estimation results of (a) Dataset 1 and (b) Dataset 2.

It is observed that under strategies 1 and 2, the ECM-based ML models generally estimate the most accurate SOH among different ML methods, even when less data is used for training. And nearly, the ECM-based GPR estimates battery SOH with the lowest errors with each ratio of split data in both Datasets 1 and 2. Under strategy 3, at all three temperatures the SVR based on ECM estimates more accurate battery SOH than other ML methods in almost all cases, indicating the high generalization capability of combing ECM and SVR on data with unseen temperatures in practice. Under strategy 4, in both datasets the ECM generally indicates the most accurate battery aging state for each ML model, and in this strategy, GPR generally performs better than other ML methods, again highlighting the superiority of the combination of GPR and ECM. Fig. 6 further justifies the high performance of using ECM to capture battery aging evolution, and the SOH prediction accuracy is improved by combining with a suitable ML method. Also, the great estimation errors under strategies 3 and 4 than those under other strategies indicate the poor 'extrapolation capability' of ML algorithms, motivating the transfer learning for improved generalization performance in this study.

**SOH estimation with shorter duration of relaxation data.** In real applications, it is not practical to put the batteries at rest for a long time (*e.g.*, 30 min or longer) after charging, which may cause inconvenience to the users.

Also, keeping the battery management system working after battery charging also causes extra energy consumption. For these reasons, it is necessary to check if effective health indicators can be extracted from shorter duration of voltage relaxation data. Fig. 7 shows the GPR-based SOH estimation results with features extracted from different lengths of relaxation data. Note that the sampling interval is 2 min in Datasets 1 and 2, and 30 s in Dataset 3. Thus, six sampling points of voltage data, which is the least amount of data required for identifying all six parameters of ECM, leads to a relaxation period of 12 min in Datasets 1 and 2, and that of 3 min in Datasets 3.

Still, the results show that as the relaxation time reduces, the ECM-based SOH estimation errors remain the lowest in most cases. And even in Dataset 3, when the relaxation time reduces to 10 min or lower, the SOH estimation errors based on ECM decline to lower values than those with other features. In Dataset 2, the estimation accuracies do not decrease as the relaxation time drops for all features, indicating a minimum relaxation time of 12 min (6 voltage data points) is sufficient for the ECM-based GPR to predicting accurate battery SOH (RMSEs within 1.1%). While for the NCA battery (Dataset 1), the estimation errors increase exponentially for all features as the relaxation time drops, and when the relaxation time drops to 16 min or lower, the estimation errors rise greatly. Therefore, the minimal relaxation time for accurate SOH estimation can be 20 min, when the estimated error based on ECM is close to 1%. In Dataset 3, the lowest estimation errors of SOH with all features lie in the relaxation time of 10-20 min, and the ECM-based GPR estimates the most accurate SOH at 10 min (relaxation time), with the RMSE less than 1.1%. Conclusively, Fig. 7 shows that for different types of batteries, the shortest relaxation times required for accurate battery SOH estimation are different, ranging from 10 min to 20 min.

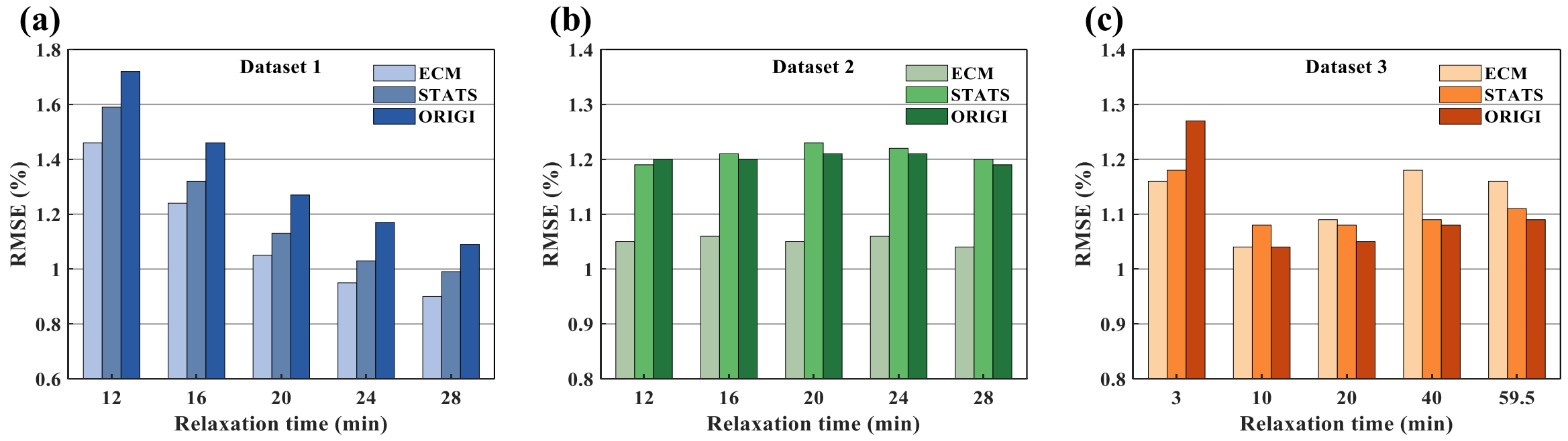

Fig. 7. SOH estimation results based on GPR with features extracted from voltage data of different relaxation times. Results of (a) Dataset 1, (b) Dataset 2, and (c) Dataset 3.

Table III lists the SOH estimation errors based on different ML methods. Still, it shows that with each ML method, the ECM-based estimated RMSE is generally the lowest among all features at each relaxation time. And the ECM-based GPR generally predicts the most accurate SOH under all circumstances, with the estimation errors close to 1% when the least required relaxation time (10 min to 20 min) is met. SVR performs slightly better than GPR in Dataset 3, with the RMSE being 0.9% at 10 min (relaxation time). However, in other datasets SVR predicts battery SOH with much larger errors than GPR, and even in Dataset 2, the estimation errors all increase as the relaxation time reduces.

Table III RMSE of SOH estimation using different ML methods with different types of features.

| Dataset | Relaxation time (min) | GPR | | | XGBoost | | | SVR | | |
|---|---|---|---|---|---|---|---|---|---|---|
| | | ECM | STATS | ORIGI | ECM | STATS | ORIGI | ECM | STATS | ORIGI |
| Dataset 1 | 12 | **1.46** | 1.59 | 1.72 | **1.58** | 1.76 | 2.04 | **2.02** | 2.03 | 2.82 |
| | 16 | **1.24** | 1.32 | 1.46 | **1.34** | 1.50 | 1.80 | **1.57** | 1.65 | 2.54 |
| | 20 | **1.05** | 1.13 | 1.27 | **1.15** | 1.34 | 1.67 | **1.22** | 1.35 | 2.16 |
| | 24 | **0.95** | 1.03 | 1.17 | **1.05** | 1.24 | 1.61 | **1.09** | 1.21 | 1.79 |
| | 28 | **0.90** | 0.99 | 1.09 | **1.02** | 1.12 | 1.50 | **1.00** | 1.11 | 1.50 |
| Dataset 2 | 12 | **1.05** | 1.19 | 1.20 | **1.26** | 1.35 | 1.48 | **1.40** | 1.50 | 2.23 |
| | 16 | **1.06** | 1.21 | 1.20 | **1.25** | 1.37 | 1.49 | **1.34** | 1.46 | 2.14 |
| | 20 | **1.05** | 1.23 | 1.21 | **1.26** | 1.36 | 1.50 | **1.31** | 1.40 | 2.10 |
| | 24 | **1.06** | 1.22 | 1.21 | **1.23** | 1.33 | 1.50 | **1.28** | 1.36 | 2.06 |
| | 28 | **1.04** | 1.20 | 1.19 | **1.19** | 1.32 | 1.48 | **1.24** | 1.31 | 2.04 |
| Dataset 3 | 3 | **1.16** | 1.18 | 1.27 | **1.16** | 1.33 | 1.94 | **1.06** | 1.15 | 1.99 |
| | 10 | **1.04** | 1.08 | **1.04** | 1.08 | **1.02** | 1.22 | **0.90** | 1.00 | 0.97 |
| | 20 | 1.09 | 1.08 | **1.05** | **1.12** | 1.23 | 1.21 | 0.95 | 0.95 | **0.92** |
| | 40 | 1.18 | 1.09 | **1.08** | 1.27 | 1.21 | **1.11** | 1.10 | **0.94** | 0.97 |
| | 59.5 | 1.16 | 1.11 | **1.09** | 1.24 | 1.19 | **1.10** | **0.88** | 1.00 | 1.03 |

*C. Battery SOH estimation with transfer learning (TL)*

Three TL methods (presented in Fig. 3) are used to evaluate the transfer learning capability of the ML methods and features. The SD data is from Dataset 1 while the TD data is from Datasets 2 and 3. To construct the TD data, only one cell is randomly selected from each operation condition, and the cycling data of the selected cells is sampled every 100 cycles. For example, for a cell with a cycle life of 600 cycles, only 6 samplings are extracted. Therefore, only limited TD data is available for TL. Fig. 8 shows the TL results of SOH estimation with GPR and ECM features. Generally, the TL results show that the SOH estimations based on all three TL methods closely captures the observations, causing the estimation errors varying around 0. The RMSEs based on TL1 are, respectively, 1.86% and 1.12% with Datasets 2 and 3, which errors are the smallest among the three TL methods. More results are presented in Fig. 9, where the TL-based SOH estimation with 20 groups of data is illustrated.

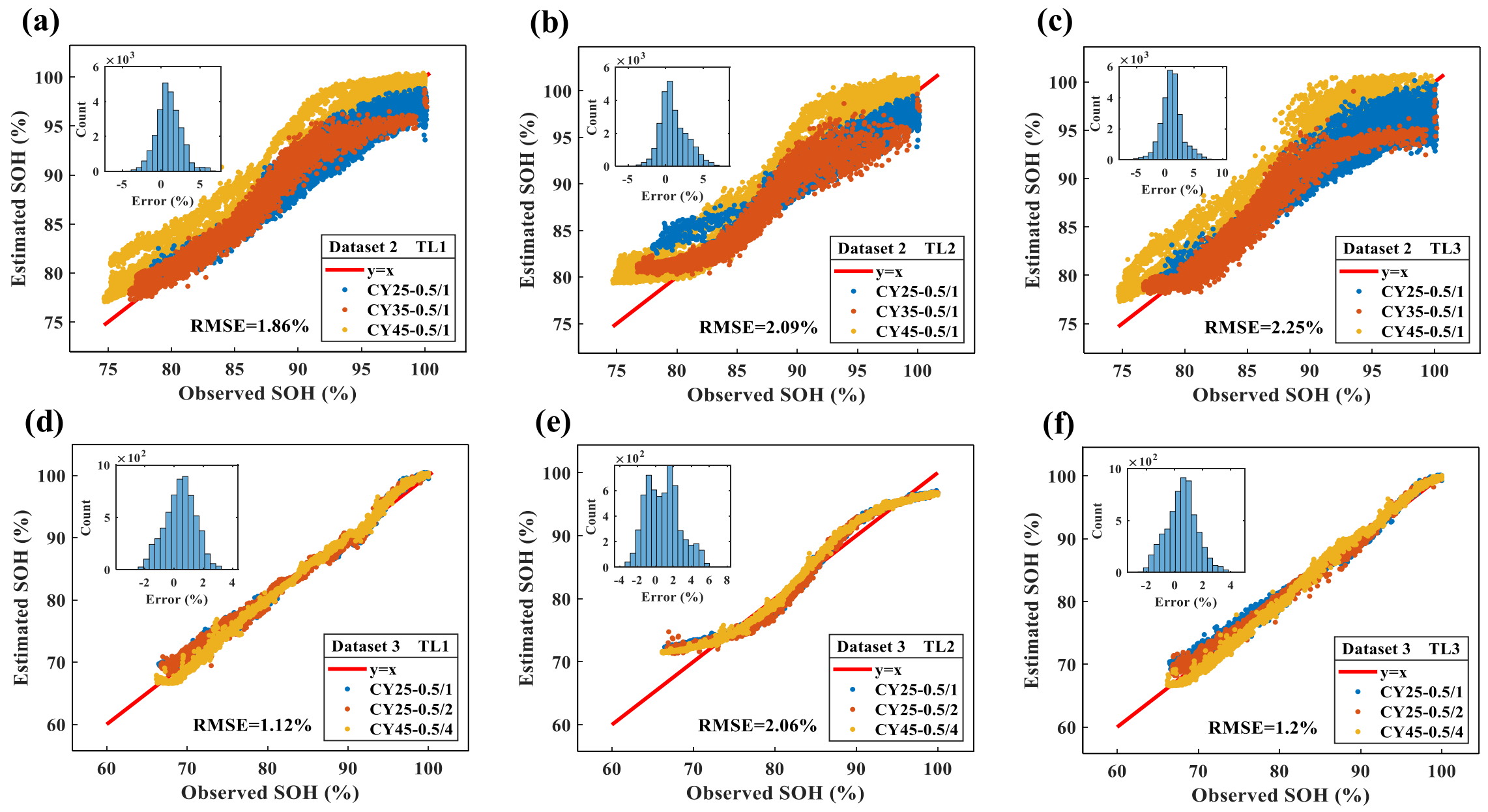

Fig. 8 TL results of SOH estimation with GPR and ECM features. Estimation results of (a) TL1, (b) TL2, and (c) TL3 on Dataset 2. Estimation results of (d) TL1, (e) TL2, and (f) TL3 on Dataset 3.

For each group of TD data, the cell from each operation condition is randomly selected. ‘No TL’ means no TL method is used, and GPR is only trained based on the limited TD data for SOH estimation. Fig. 9 shows that the TL performance from Dataset 1 to Dataset 2 is better than that from Dataset 1 to Dataset 3, which might be caused by the quite different aging characteristics of Dataset 3 from those of both Datasets 1 and 2 (Fig. 4). With Dataset 2, all three TL methods work effectively with more accurate SOH estimated than ‘No TL’. With Dataset 3, only TL1 and TL3 performs well, while TL2 estimates SOH with larger errors than ‘No TL’. Fig. 9 shows that TL1 generally estimates the most accurate SOH with both datasets and different ML methods, with the SOH estimation errors significantly reduced compared with those of ‘No TL’. The averaged RMSE in each case is listed in Table IV, where the TL results based on other ML models and features are also shown.

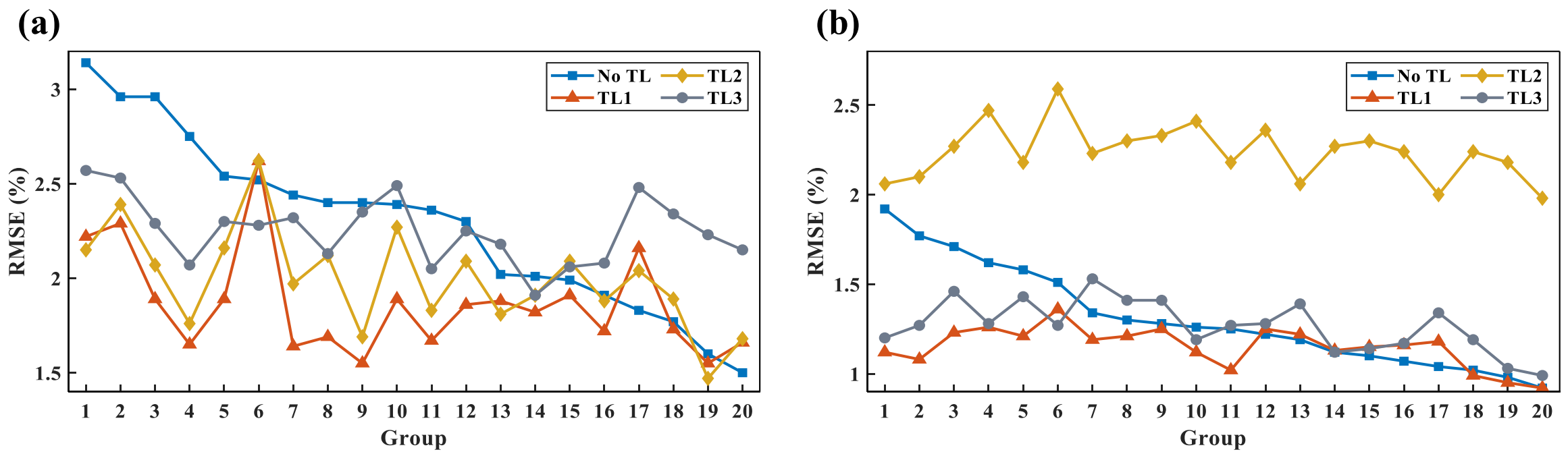

Fig. 9 TL results based on 20 groups of randomly constructed TD data. SOH estimation results on (a) Dataset 2, (b) Dataset 3.

‘ZSL’ indicates the zero-shot learning, in which case the ML model is only trained based on SD data and the trained model is then used to estimate battery SOH on the test data directly. Table IV shows that ZSL predicts battery SOH with large errors, and these estimation errors are further enlarged with Dataset 3, owing to the greatly deviated ECM features of Dataset 3 from those of Dataset 1 (SD data). These large estimation errors cause the necessity to introduce limited TD data in TL for improved prediction performance. For Dataset 2 with strong cell-to-cell variations of aging characteristics, which is usually the case in practice, TL1 with ECM features generally predicts the most accurate SOH against different ML methods. And when GPR is used, the ECM-based TL1 estimates SOH with the smallest RMSE (1.87%), with the accuracy improved by about 20% than that of ‘No TL’ (RMSE of 2.29%).

For Dataset 3, owing to the high cell-to-cell uniformities of battery aging, in most cases ‘No TL’, when only limited TD data is used for training, obtains the most accurate SOH estimation against different ML methods. However, TL1 based on GPR and ECM still predicts accurate battery SOH with an RMSE being 1.15%, which is second small to the estimation error of 1.09% that is obtained by ‘No TL’ based on GPR. Conclusively, the TL1 method, in which case the training data is augmented by directly introducing the TD data, generally obtains the most accurate battery SOH based on GPR and ECM, again validating the excellence of ECM features and the improved prediction capability with GPR.

Table IV RMSE of TL results based on different ML methods and features.

| Dataset | Method | GPR | | | XGBoost | | | SVR | | |
|---|---|---|---|---|---|---|---|---|---|---|
| | | ECM | STATS | ORIGI | ECM | STATS | ORIGI | ECM | STATS | ORIGI |
| Dataset 2 | ZSL | 2.18 | 2.41 | 2.72 | 2.48 | 2.88 | 3.60 | 2.36 | 2.93 | 3.58 |
| | No TL | 2.29 | 2.17 | 2.91 | 2.79 | 2.93 | 3.31 | 2.52 | 2.44 | 3.25 |
| | TL1 | **1.87** | **2.14** | 2.49 | **2.13** | **2.38** | **3.09** | **2.17** | **2.38** | 3.04 |
| | TL2 | 1.99 | 2.30 | **2.18** | 2.92 | 3.01 | 3.36 | **2.17** | 2.63 | **2.83** |
| | TL3 | 2.25 | 2.60 | 3.01 | 2.32 | 3.07 | 3.28 | 2.22 | 2.48 | 3.17 |
| Dataset 3 | ZSL | 4.76 | 17.05 | 4.63 | 4.74 | 10.39 | 5.05 | 14.64 | 17.57 | 7.35 |
| | No TL | 1.31 | 1.27 | **1.09** | **1.68** | **1.70** | **1.97** | **1.34** | **1.46** | 1.60 |
| | TL1 | **1.15** | 1.34 | 1.67 | 1.77 | 1.80 | 2.36 | 3.35 | 4.20 | 3.37 |
| | TL2 | 2.24 | 2.87 | 1.41 | 3.10 | 3.30 | 2.19 | 2.37 | 2.20 | **1.40** |
| | TL3 | 1.27 | **1.17** | 1.32 | 1.86 | 2.10 | 2.50 | 1.49 | 2.98 | 3.00 |

## V. CONCLUSIONS

Battery state of health (SOH), as an important indicator to battery aging state, is the basis to define battery aging failure. Accurately estimating battery SOH is a core function of battery management system (BMS), based on which in-time maintenance of batteries is informed and conducted to ensure them operating properly and safely.

This study develops a physics-informed machine learning (ML) methodology for battery SOH estimation, where physical features with high interpretability are used to indicate battery aging evolution. The physics-informed features are extracted from an equivalent circuit model (ECM). Besides, other two typical types of features including the statistical features (namely STATS) and the original features (namely ORIGI) are also used for systematically comparisons. For fair comparisons, all features are extracted from the voltage relaxation data after battery full charging, which data can be easily collected through BMS in practice. The Gaussian process regression (GPR) is used to learn the relationships between the input features and battery SOH, while other two advanced ML methods including the extreme gradient boosting (XGBoost) and the support vector regression (SVR) are also introduced for comparison purposes.

The developed methodology is used to estimate battery SOH under different scenarios. The voltage data at different relaxation times is used for features extraction to evaluate the feasibility of the methodology. The robustness of the ML models combined with three types of features is checked against different ratios of training/test data. And the generalization capability of the data-driven methods is validated against test data at unseen temperatures and aging states. The transfer learning (TL) capability is verified by developing three TL methods including ‘Augmentation’ (TL1), ‘Features transfer’ (TL2) and ‘Delta learning’ (TL3). The method performance under these scenarios are validated using the cycling data of three types of batteries with a total number of 118 cells and 58826 data units.

Experimental results show that the minimal relaxed time required for accurate battery SOH estimation is between 10 to 20 min, when the ECM-based GPR estimates SOH with the root mean squared errors (RMSEs) close to 1%. The ML models based on different features are robust against different ratios of training/test data, and still, the GPR based on ECM performs the best. The ECM features exhibit excellent generalization capability of indicating battery SOH, which is validated by the test data under unseen operation conditions. And TL1 based on GPR and ECM transfers the knowledge learned from the source domain data more effectively, with the estimation accuracy improved by approximately 20% compared with that of ‘No TL’. In summary, the results highlight that the physical features of ECM generally indicate battery aging evolution more reliably and accurately than other features. And the GPR model with ECM features as input estimates the most accurate battery SOH under most scenarios.

## APPENDIX

### STATISTICAL FEATURES

The six STATS features are calculated as follows:

$$Max = max\{x_i\}, \tag{15}$$

$$Mean = \frac{1}{n}\sum_{i=1}^{n} x_i, \tag{16}$$

$$Min = min\{x_i\}, \tag{17}$$

$$Var = \frac{1}{n-1}\sum_{i=1}^{n}(x_i - \bar{x})^2, \quad (18)$$

$$Ske = \frac{\frac{1}{n}\sum_{i=1}^{n}(x_i - \bar{x})^3}{\left(\frac{1}{n}\sum_{i=1}^{n}(x_i - \bar{x})^2\right)^{\frac{3}{2}}}, \quad (19)$$

$$Kur = \frac{\frac{1}{n}\sum_{i=1}^{n}(x_i - \bar{x})^4}{\left(\frac{1}{n}\sum_{i=1}^{n}(x_i - \bar{x})^2\right)^{2}} - 3, \quad (20)$$

where $x_i$ is the $i$th sampling data, and $n$ is the total number of the data points. $Max$, $Mean$, $Min$, $Var$, $Ske$, and $Kur$ indicate the maxima, mean, minima, variance, skewness, and the excess kurtosis of the sampling data, respectively.

## Acknowledgement

This work was partially supported by the National Key Technology Research and Development Program of China (No. 2021YFB2402002).

## Author contributions

Xinhong Feng: Methodology, Modeling, Writing-original draft.
Yongzhi Zhang: Conceptualization, Supervision, Methodology, Writing-original draft, reviewing&editing.
Rui Xiong: Supervision, Writing-Reviewing&Editing.
Chun Wang: Writing-Reviewing&Editing.

## Competing interests

The authors declare no conflicts of interets.